\documentclass[preprint,12pt]{elsarticle}

\usepackage{easyReview}
\usepackage[normalem]{ulem}
\usepackage[left=2cm,right=2cm,top=2cm,bottom=2cm]{geometry}
\usepackage{caption}
\usepackage{subcaption}

\usepackage{amsmath}

\begin{document}
	
\begin{frontmatter}
	
\title{Transport with noise in dilute gases: Effect of Langevin thermostat on transport coefficients.}

\author{A. Alés, J. I. Cerato, L. Marchioni and M. Hoyuelos}
\affiliation{organization={Instituto de Investigaciones Físicas de Mar del Plata (IFIMAR-CONICET), Departamento de Física, Facultad de Ciencias Exactas y Naturales, Universidad Nacional de Mar del Plata(NM},
	addressline={Funes 3350}, 
	city={Mar del Plata},
	postcode={7600}, 
	country={Argentina}}

\begin{abstract}
In dilute gases, transport properties such as the thermal conductivity, self-diffusion, and viscosity are significantly affected by interatomic collisions, which are determined by the potential form. This study explores these transport properties in the presence of a Langevin thermostat in systems where particles interact through various potentials, including soft-core and hard-core potentials, both with and without an attractive region. Using molecular dynamics simulations and a theoretical approach based on an analogy with an electric circuit (Ohm's law), we derived and compared the transport coefficients across these interatomic potentials for different couplings with the thermostat. The transport coefficients were obtained by considering the thermostat as a resistance in a series circuit.
\end{abstract}

\end{frontmatter}

\section{Introduction}
\label{sc:in}

Boltzmann's kinetic theory of gases is essential for understanding the transport properties of dilute gases \cite{hansen2013theory}. Within the framework established by Chapman and Enskog \cite{chapman1990mathematical,hirschfelder1964molecular}, this theory provides the basis for deriving the coefficients of thermal conductivity, self-diffusivity, and viscosity that describe the transport of energy, mass, and momentum, respectively. These coefficients relate microscopic properties, such as the size or mass of the particles, with macroscopic behaviors. This link is fundamental for interpreting the behavior of gases across various fields, from chemical engineering to astrophysics. 

In the Boltzmann theory, particle collisions are described through interatomic potentials \cite{hirschfelder1964molecular}. For example, the hard-sphere potential, which takes a zero value beyond a certain radius and infinity within it, is widely used in numerical simulations as a preliminary step in analyzing systems with more complex interactions \cite{heyes2007self,jover2012pseudo}. In general, interatomic potentials typically consist of a repulsive component at short distances, and may include one or more attractive or repulsive contributions at intermediate ranges, eventually decaying to zero at long distances. In his pioneering work, Lennard-Jones introduced a potential as a function of the interatomic distance $r$, with an attractive term proportional to $r^{-6}$, derived from consideration of dipole moment fluctuations \cite{lennard1931cohesion}. For convenience, the repulsive term is often modeled as $r^{-12}$, although numerous studies have explored alternative power law dependencies. Hard-core potentials, such as the Lennard-Jones potential, impose an infinite kinetic energy requirement for particles to approach each other, whereas soft-core potentials maintain a finite repulsive energy at short distances, such as the Morse potential \cite{bahethi1964properties,saxena1964transport}. Notably, the presence of attractive regions in a potential opens a range of properties in gas behavior. When there is an attractive depth-well potential interaction, particles can exhibit binding or orbiting phenomena if they collide with specific impact parameters and energies \cite{hirschfelder1964molecular,mason1988transport}, adding further complexity to the analysis of gas dynamics. 

The measurement of transport coefficients for these potentials in dilute gases has been a well-established practice since the early days of computational simulations. However, the inclusion of a thermal bath introduces significant modifications to the transport coefficients because its influence disrupts the natural dynamics of the system. Historically, it has been recommended to measure these coefficients only when the thermal bath provides a very long damping time, thus minimizing its impact on the intrinsic transport properties \cite{tanaka1983constant}; see also Sec.\ 6.1.1 in \cite{frenkel}.

This external influence from the thermostat was initially thought to obscure the essence of the transport coefficient measurements. For example, the Andersen thermal bath substantially alters the transport coefficient values, unless the damping time is sufficiently large. Despite these concerns, recent studies have demonstrated that for viscosity \cite{marchioni2024viscosity} and self-diffusion coefficients \cite{marchioni2023dependence}, the effect of a Langevin thermal bath is predictable and quantifiable. Specifically, this influence can be attributed to the coupling of two independent transport contributions: one arising from interatomic interactions and the other arising from the thermal bath in which the particles are immersed. For brevity, in the following we use ``diffusion'' instead of ``self-diffusion'' to refer to diffusion of tagged particles in a medium composed by particles of the same kind.

Moreover, the damping factor of the Langevin thermostat serves as a  control parameter \cite{marchioni2023dependence,marchioni2024viscosity}. By tuning this factor, one can transition from a regime dominated by the thermal bath, characterized by small damping times, to a regime where the influence of the bath becomes negligible as the damping time approaches infinity. 

The aim of this paper is to study, both numerically through molecular dynamics simulations and analytically using generalized Ohm's laws, the behavior of the transport coefficients for hard- and soft-core potentials, with or without an attractive part, in the presence of a Langevin thermostat. The behavior of the transport coefficients as functions of the damping time is obtained considering that the effect of the thermostat can be represented as a resistance in a series circuit.

This paper is structured as follows. The methodology used is detailed in Section \ref{sc:mt}, the analytical and computational results are presented in Section \ref{sc:ra}, and conclusions are provided in Section \ref{sc:cc}.

\section{Methodology}
\label{sc:mt}

Near equilibrium, every macroscopic transport process is linearly related to the correlations of microscopic fluctuations around the equilibrium state. The Green-Kubo relations provide a rigorous framework to obtain these coefficients by expressing them as integrals over time-correlation functions of microscopic currents in equilibrium systems without requiring the application of external gradients \cite{kubo,reichl2016modern}. However, while these relations establish a connection between transport coefficients and equilibrium fluctuations, they do not provide explicit expressions for the correlation functions themselves. In contrast, the Chapman-Enskog method, based on the Boltzmann equation, allows for explicit computation of these correlations in the case of dilute gases by employing a perturbative expansion of the distribution function \cite{chapman1990mathematical}. 

The Langevin thermostat introduces an internal stochastic current, which we refer to as the Langevin current. This current arises from the random forces exerted by the thermostat, which mimic the thermal fluctuations experienced by particles in a fluid. These fluctuations follow the statistical principles of Brownian motion, resulting in a dynamic that contributes to the overall transport behavior. 

We present the simulation details in subsection \ref{ss:md} and briefly review the calculation of the transport coefficients using the Chapman-Enksog approach for a particle system without a thermal bath. For a dilute gas in a thermal bath, if interactions are neglected and particle dynamics is essentially Brownian, transport coefficients are obtained by means of Green-Kubo calculations; see subsection \ref{ss:tb}.

\subsection{Simulation Details}
\label{ss:md}

\begin{figure}[t!]
	\begin{center}		
		\includegraphics[width=0.65\columnwidth]{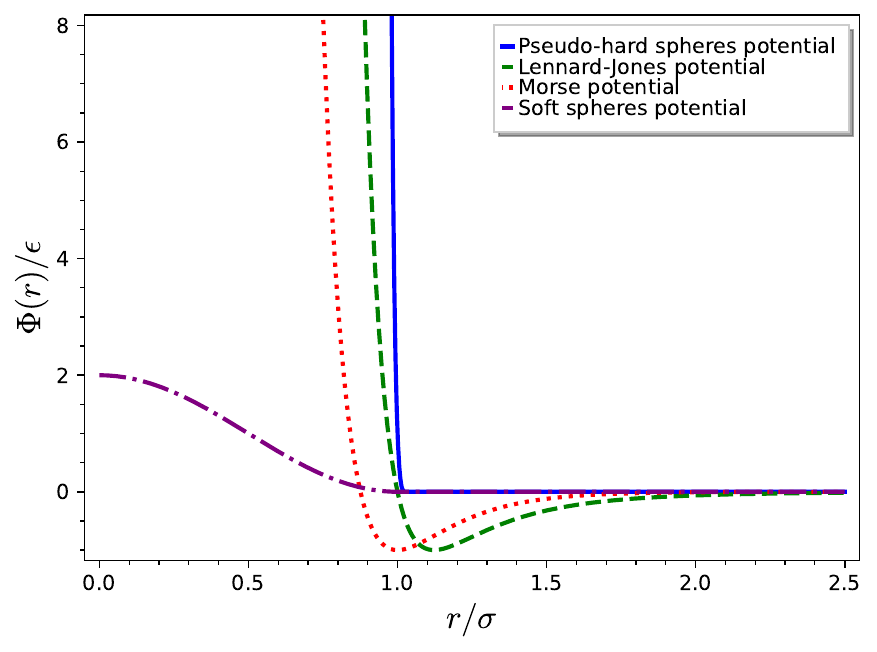}
		\caption{Scaled interatomic potentials used in this paper, $\Phi/\epsilon$, against distance, $r/\sigma$ ($\epsilon$ and $\sigma$ are characteristic values of energy and distance for each potential). The potentials are pesudo hard spheres (blue line), Lennard-Jones (LJ, green dashed curve), Morse (red dotted curve) and soft spheres (purple dot-dashed curve). Soft spheres and Morse potentials are soft core, while LJ and pseudo hard spheres potentials are hard core (they diverge as $r\rightarrow 0$).}
		\label{fig:Potentials}
	\end{center}
\end{figure}

Molecular dynamic simulations were implemented using LAMMPS software \cite{plimpton} (Large-scale Atomic-Molecular Massively Parallel Simulator). Each interatomic potential has characteristic length $\sigma_\mu$ and energy $\epsilon_\mu$, where the subindex $\mu$ identifies the potential.  The particle density is $\rho=N/V$, where $N$ is the particle number, $V$ is the system volume, and the reduced dimensionless particle density is given by $\rho^* = \rho\, \sigma_\mu^3$. The reduced temperature $T^*$ in terms of the temperature $T$ is $T^* = k_B T/\epsilon_\mu$. In all cases, we used a small reduced density of $\rho^* = 0.01$ for which Boltzmann theory approximately holds. To represent qualitatively different interactions, we select the following four interatomic potentials $\Phi_\mu(r)$: 

\begin{enumerate}
    \item Pseudo-hard spheres: 
    \begin{equation*}
    \Phi_{\rm HS}(r) = \left\{ \begin{array}{cl}
          \frac{\epsilon_{\rm HS}\, \lambda_r}{\lambda_r-\lambda_a} \left( \frac{\lambda_r}{\lambda_a} \right)^{\lambda_a} \left[ \left( \frac{\sigma_{\rm HS}}{r}\right)^{\lambda_r} - \left( \frac{\sigma_{\rm HS}}{r}\right)^{\lambda_a}\right] + \epsilon_{\rm HS} &  \mbox{for } r < \sigma_{\rm HS} (\frac{\lambda_r}{\lambda_a})^{\frac{1}{\lambda_r-\lambda_a}}\\
         0 &  \mbox{for } r \ge \sigma_{\rm HS} (\frac{\lambda_r}{\lambda_a})^{\frac{1}{\lambda_r-\lambda_a}}
    \end{array} \right. 
    \end{equation*}
    where the scheme of WCA (Weeks-Chandler-Andersen) \cite{weeks1971role} is used. This potential correctly reproduces the hard-sphere properties for $\lambda_r = 50$ and $\lambda_a =49$, with a reduced temperature $T^* = 1.5$, as shown in Ref.\ \cite{jover2012pseudo}.    In the simulations, we used $\epsilon_{\rm HS} = 1$ and $\sigma_{\rm HS}=1$ in LJ units.
    \item Lennard-Jones (LJ) is the most common interatomic potential for simple fluid modeling: $$\Phi_{\rm LJ}(r)  = \epsilon_{\rm LJ} \left[ \left( \frac{\sigma_{\rm LJ}}{r}\right)^{12} - \left( \frac{\sigma_{\rm LJ}}{r}\right)^{6}\right].$$ 
    A cutoff radius $r_c = 2.5$ was used, and the values $\epsilon_{\rm LJ}=1$ and $\sigma_{\rm LJ}=1$ in LJ units were used.
    \item Morse potential, given by $$ \Phi_{\rm M}(r) = \epsilon_{\rm M}  \left ( e^{- 2\alpha (r - \sigma_{\rm M}) } -   2  e^{- \alpha (r - \sigma_{\rm M}) } \right)$$ 
    (see \cite{smith1964automatic}). The parameters of this potential has been chosen to fit the well of the Lennard-Jones region for direct comparison, and the values are  $\epsilon_{\rm M} = 1.10\, \epsilon_{\rm LJ}$, $\alpha =  5.58/\sigma_{\rm LJ} $ and $\sigma_{\rm M} = 1.12\,\sigma_{\rm LJ}$. 
    \item At last, soft-spheres interatomic potential, modeled as 
    \begin{equation*}
        \Phi_{\rm SS}(r) = \left\{ \begin{array}{cc}
          \epsilon_{\rm SS} \left[1 + \cos( \pi r /\sigma_{\rm SS}) \right]  & \mbox{for } r < \sigma_{\rm SS} \\
           0  & \mbox{for } r \ge \sigma_{\rm SS}
        \end{array}  \right.,
    \end{equation*}
     where we used $\epsilon_{\rm SS} = 50$ and $\sigma_{\rm SS}=1$ in LJ units for the simulations.   
\end{enumerate} 

The shapes of these potentials are shown in Fig.\ \ref{fig:Potentials}. Whereas the first two are hard-core potentials (they diverge for $r\rightarrow 0$), the last two are soft-core potentials. The pseudo-hard and soft spheres are potentials with only a repulsive part, whereas the remaining potentials have an attractive depth well. The temperature used for the pseudo-hard sphere potential is $T^*=1.5$ (that, as mentioned before, is the temperature for which hard-sphere properties are reproduced) and for soft-spheres is $T^*=4$. Lennard-Jones and Morse potentials are studied considering both temperatures.

Molecular dynamics (MD) simulations were performed with 32,000 particles, including two equilibration stages. In the first stage, Gaussian-distributed velocities were assigned to the particles, which were initially arranged in an fcc lattice, and the system was evolved in the NVT ensemble using a Langevin thermostat with a specified damping time $t_d$. After $2\times10^5$ time steps, the simulation was switched to the NVE ensemble and continued for the same number of steps to allow the system to equilibrate. Subsequently, a data-gathering run of $1.5\times10^6$ steps was carried out in the NVE ensemble. A time step of 0.001 was employed in all cases, except for the pseudo-hard sphere potential, for which a smaller time step of 0.0001 was required to ensure the dynamical stability of the system.

\subsection{Theoretical Background}
\label{ss:tb}

The Boltzmann equation holds for a dilute gas; it is an integro-differential equation that describes the one-particle nonequilibrium distribution function, $f(\mathbf{r},\mathbf{v},t)$, which gives the average number of particles with position $\mathbf{r}$ and velocity $\mathbf{v}$ at time $t$ \cite{reichl2016modern,mcquarrie}.  The Chapman-Enskog method uses this distribution function to determine transport coefficients in the zero-density limit. A perturbative expansion is used, where $f(\mathbf{r},\mathbf{v},t)$ is expressed as a series around the equilibrium distribution of the Boltzmann equation. The results align with macroscopic hydrodynamic equations by stopping the expansion after the first correction \cite{chapman1990mathematical,mcquarrie}. The transport coefficients are then calculated as a series involving Sonine polynomials, which typically converge quickly, meaning that only a few terms are needed for accurate results \cite{hirschfelder1964molecular}. 
The transport coefficients for dilute gases are
\begin{equation}
    \lambda_B = \frac{75\, k_B}{64\, \sigma^2} \left(\frac{k_B T}{\pi  m} \right)^{1/2} \frac{ f_\lambda}{\Omega^{(2,2)}},
    \label{eq:LambdaB}
\end{equation}
for thermal conductivity, where $m$ is the particle mass, $T$ is temperature and $k_B$ is Boltzmann's constant; we have the following expression for the diffusion coefficient,  
\begin{equation}
    D_B = \frac{3}{8\, \rho\, \sigma^2 } \left(\frac{ k_B T}{ m \pi } \right)^{1/2} \frac{ f_D}{\Omega^{(1,1)}};
    \label{eq:DiffB}
\end{equation}
and the viscosity coefficient is
\begin{equation}
    \eta_B = \frac{5}{16\, \sigma^2} \left(\frac{m k_B T}{\pi} \right)^{1/2}  \frac{ f_\nu}{\Omega^{(2,2)}},
    \label{eq:nuB}
\end{equation}
where $\Omega^{(i,j)}$ are collision integrals relative to the hard sphere system, which depend on the temperature and the type of interatomic potential considered \cite{hirschfelder1964molecular,mcquarrie}. We have included the subscript $B$ to indicate that these coefficients were derived by considering particle collisions within the framework of the Boltzmann equation \cite{reichl2016modern,mcquarrie}. The $f_i$ functions, known as correction factors, are expressed in terms of the collision integrals \cite{mason1988transport,monchick1961transport}. Since it is well known that deviations of $f_i$ from unity are very small, we consider $f_i = 1$, with $i = \nu,D,\lambda$; see \cite{viehland1995high} or Ch.\ 19 in \cite{mcquarrie}.

The dimensionless coefficients are obtained using the following scaling: $D^* = D \,\sigma^{-1}\sqrt{m/\epsilon}$, $\eta^* = \eta\, \sigma^2/\sqrt{m\epsilon}$ and $\lambda^* = \lambda\,k_B^{-1}\sigma^2 \sqrt{m/\epsilon}$ (subindex $\mu$ in $\sigma$ and $\epsilon$ is omitted for simplicity). By applying the scaling to Eqs.\ \eqref{eq:DiffB}, \eqref{eq:nuB} and \eqref{eq:LambdaB} we obtain:
\begin{align}
    D_B^* &= \frac{3}{8 \,\rho^*} \frac{\sqrt{T^*/\pi}}{\Omega^{*(1,1)}}, \label{e.DBa} \\
    \eta_B^* &= \frac{5}{16} \frac{\sqrt{T^*/\pi}}{\Omega^{*(2,2)}}, \label{e.eBa} \\
    \lambda_B^* &= \frac{75}{64} \frac{\sqrt{T^*/\pi}}{\Omega^{*(2,2)}}. \label{e.lBa}
\end{align}

From another perspective, the Green-Kubo relations provide a framework to obtain these coefficients by relating them to the equilibrium time correlations of microscopic currents \cite{chapman1990mathematical,reichl2016modern,kubo}. These relationships are particularly powerful because they allow the calculation of transport properties in equilibrium systems without requiring the application of external gradients. The correlations are commonly obtained using numerical methods; however, as explained in subsection \ref{ss:te}, analytical results are possible for the Brownian dynamics produced by the Langevin thermostat for small concentrations. For example, the thermal conductivity $\lambda$ is expressed as
\begin{equation}
    \lambda = \frac{1}{k_B T^2 V} \int_0^{\infty} \langle J_Q(0) J_Q(t) \rangle \, dt,
\end{equation}
where $J_Q(t)$ is the energy current in an arbitrary direction (the system is isotropic), $V$ is the system volume, and $ \langle J_Q(0) J_Q(t) \rangle$ denotes the ensemble average of the correlation between the current at time zero and time $t$ (more details on this topic in Sec.\ \ref{ss:te}). 

We aimed to analyze a system with atomic collisions in a thermal bath at a low concentration. A binary gas mixture can be considered for this description. The system consists of the particles of interest, at low density, of mass $m$; they are immersed in a thermal bath represented by particles with mass $m'$, with $m \gg m'$. Interatomic collisions are as follows: $m$-$m$ collisions, governed by the chosen interatomic potential, are rare due to the low particle density; $m$-$m'$ collisions, on the other hand, are more frequent and follow the well-established rules of Brownian motion. Collisions $m'$-$m'$ among thermal bath particles are not relevant in this scenario. This picture indicates that the transport of any conserved quantity has to overcome the resistance produced by both types of collisions, $m$-$m$ and $m$-$m'$, as in a series circuit.

Generalized Ohm's laws are linear relations between thermodynamic forces (gradients of chemical potential, temperature or velocity) and currents (of particles, heat or momentum); the proportionality constants are the transport coefficients. These laws are fundamental to the development of hydrodynamic equations in classical irreversible thermodynamics. The entropy production in the hydrodynamic equations can be written as the product of currents times forces, analogous to the Joule heat. Here, we consider the consequences of this analogy when resistances in series are present. For example, let us consider the current of tagged particles. The presence of other particles of the same type causes resistance to the particle current. In the presence of the thermostat, there is also the resistance caused by the small particles of the reservoir, whose effect is represented by random forces and damping. The particle current has to overcome both resistances, therefore, they are in series. The same reasoning can be applied to heat or momentum currents. Transport coefficients represent, instead of resistances, the conductances. We call $\mathcal{C}_B$ the conductance in the absence of the thermostat that is obtained from the Boltzmann's theory, and $\mathcal{C}_L$ the conductance introduced by the Langevin thermostat. Since the circuit is in series, the resulting conductance is

\begin{equation}
	\frac{1}{\mathcal{C}} = \frac{1}{\mathcal{C}_L} + \frac{1}{\mathcal{C}_B}.
	\label{eq:rg}
\end{equation}

This concept, which extends the consequences of the electric circuit analogy in hydrodynamic equations, was introduced in  \cite{marchioni2023dependence,marchioni2024viscosity}. Depending on the damping time of the Langevin thermostat, the transport coefficients can be dominated by one or the other contribution. If the damping time is short, the system responds according to the Brownian contribution to the conductance, whereas for long damping times, it behaves according to the Boltzmann contribution.

\section{Results and Analysis}
\label{sc:ra}

\subsection{Transport coefficient calculation for Langevin Currents}
\label{ss:te}

Considering an arbitrary direction in an isotropic system, the Langevin thermostat produces a force, on particle $i$, that is given by 
\begin{equation}
F_i (t) = - m v_i / t_d + \alpha_i(t),
\end{equation}
where $m$ and $v_i$ are the mass and velocity of particle $i$, $t_d$ is the damping time, and  $\alpha_i(s)$ is a random force with zero mean, $\langle\alpha_i(t)\rangle=0$, and delta time correlated, $\langle \alpha_i(t) \alpha_i(t') \rangle = \frac{2 m k_B T}{t_d} \delta(t-t')$; the noise amplitude is determined by the fluctuation dissipation theorem. Force $F_i$, velocity $v_i$ and random force $\alpha_i$ correspond to the components in an arbitrary direction of vectors $\mathbf{F}_i$, $\mathbf{v}_i$ and $\boldsymbol{\alpha}_i$. Under this conditions, the diffusion coefficient is well known (see, for example, \cite{reichl2016modern}); it is given by the Einstein relation,
\begin{equation}
	D_L = \frac{k_B T\, t_d}{m}.
\end{equation}
The viscosity coefficient of the system composed of Brownian particles (not to be confused with the viscosity of the medium) is (see \cite{marchioni2024viscosity})
\begin{equation}
	\eta_L = \frac{\rho\, k_B T\, t_d}{2}.
\end{equation}

For thermal conductivity, we take the following approach. In the absence of both interactions between particles and external potentials, the energy flux for an $N$ particle system in a given direction is $J_Q(t) =\sum_{i=1}^{N} \left ( \frac{m}{2} \mathbf{v}_i(t)^2 - \langle h_i  \rangle \right) v_i(t)$, where $\langle h_i  \rangle$  is the mean enthalpy per particle, that takes the value $\frac{5}{2} k_B T$ for an ideal gas (see, for example, Sec. 21.8 in \cite{mcquarrie}). Using the Green-Kubo expression for thermal conductivity, we have, 
\begin{align}
    \lambda_L =& \frac{1}{V k_B T^2} \int_0^\infty \langle J_Q(t) J_Q(0) \rangle \nonumber \\
     =& \frac{1}{V k_B T^2}  \int_0^\infty 
    \left\langle \sum_{i,j} \left( \frac{m}{2} \mathbf{v}_i(t)^2 - \langle h_i  \rangle \right) v_i(t) \left( \frac{m}{2} \mathbf{v}_j(0)^2 - \langle h_j  \rangle \right) v_j(0) \right\rangle  dt \nonumber\\
     =& \frac{\rho}{k_B T^2}  \int_0^\infty 
    \left\langle \left( \frac{m}{2} \mathbf{v}_i(t)^2 - \frac{5}{2} k_B T \right) v_i(t) \left( \frac{m}{2} \mathbf{v}_i(0)^2 - \frac{5}{2} k_B T \right) v_i(0) \right\rangle  dt, \label{e.lL}
\end{align}
where correlations between different Brownian particles are neglected; in the last line, $N/V$ is replaced by $\rho$ and subindex $i$ corresponds to any particle (symmetry under exchange of particle labels is assumed). Now, we use the expression for the velocity of a Brownian particle (see, for example, Sec. 7.2.1 in \cite{reichl2016modern}),
\begin{equation}
	v_i(t) = v_i(0) + \int_0^t ds e^{-\gamma(t-s)} \alpha_i(s).   \label{eq:corrvel}
\end{equation}
Replacing \eqref{eq:corrvel} in \eqref{e.lL}, and using the statistical properties of the random force $\alpha_i$, the integral can be solved to obtain,
\begin{equation}
    \lambda_L =  \frac{ \rho\, k_B^2\, T\, t_d}{2 \, m}.
    \label{eq:LambdaE}
\end{equation}

The dimensionless coefficients, then, become
\begin{align}
    D_L^* &= T^* t_d^* \label{e.DLa} \\
    \eta_L^* &= \rho^* T^* t_d^*/2 \label{e.eLa}\\
    \lambda_L^* &= \rho^* T^* t_d^*/2,\label{e.lLa}
\end{align}
where $t_d^* = t_d\, \sigma^{-1} \sqrt{\epsilon/m}$. The Einstein relation for the diffusion coefficient, $D_L$, has fundamental historical and experimental importance in understanding Brownian motion. The expressions for the viscosity and thermal conductivity, $\eta_L$ and $\lambda_L$, are interesting from a theoretical and numerical point of view, but it is probably not possible to verify them experimentally because of the difficulties in separating the transport of heat or momentum of the particle system from the transport of the same quantities that occur in the reservoir where the system is embedded.

\subsection{Calculation of the transport coefficients with noise}
\label{ss:calc}

In Eq.\ \eqref{eq:rg} any transport coefficient is represented by the conductance $\mathcal{C}$. The dimensionless expressions for the coefficients from Boltzmann theory, Eqs.\ \eqref{e.DBa}-\eqref{e.lBa}, and from the Langevin theory, Eqs.\ \eqref{e.DLa}-\eqref{e.lLa}, are used to calculate, in Eq.\ \eqref{eq:rg}, the transport coefficients at small concentration in the presence of the Langevin thermostat assuming a series circuit. The results are,
\begin{align}
    D^* &= \frac{T^*\, t_d^*}{\frac{8}{3} \rho^*\, t_d^*\, \Omega^{*(1,1)} \sqrt{\pi\, T^*} + 1}, \label{e.D} \\
    \eta^* &= \frac{T^*\, t_d^*\, \rho^*}{\frac{16}{5} \rho^*\, t_d^* \, \Omega^{*(2,2)} \sqrt{\pi\, T^*} + 2}, \label{e.e} \\
    \lambda^* &= \frac{T^*\, t_d^*\, \rho^*}{\frac{64}{75} \rho^*\, t_d^* \, \Omega^{*(2,2)} \sqrt{\pi\, T^*} + 2}.\label{e.l}
\end{align}
It can be seen that, for large $t_d^*$, we recover the expressions from the Boltzmann theory, Eqs.\ \eqref{e.DBa}-\eqref{e.lBa}, while for small $t_d^*$ we have the Langevin results, Eqs.\ \eqref{e.DLa}-\eqref{e.lLa}. Considering that the collision integrals are of order 1, a large or small $t_d^*$ means $t_d^* \gg 1/(\rho^* \, \sqrt{T^*})$ or $t_d^* \ll 1/(\rho^* \, \sqrt{T^*})$. We obtained numerical results, presented in the next section, that verified these expressions as functions of damping time, $t_d^*$, for fixed values of temperature and concentration ($\rho^*=0.01$).

Collision integrals can be calculated using the following nested integrals \cite{hirschfelder1964molecular},
\begin{equation}
    \Omega^{(l,s)}(T) = \left( \frac{k_B T}{2 \pi m} \right)^{1/2} \int_0^\infty dx\; e^{-x^2} x^{(2\,s +3)} S^l (k_B T x),
    \label{eq:IntCol1}
\end{equation}
where $S^l (k_B T x)$ is the collision cross section, a function of the initial kinetic energy of colliding particles; is is defined as
\begin{equation}
    S^{l}(E) =  2\, \pi \int_0^\infty b \left[ 1 - \cos^l\left( \chi(b,E) \right ) \right] db.
    \label{eq:IntCol2}
\end{equation}
Here, $b$ denotes the impact parameter and $\chi$ the classical deflection angle as a function of the relative kinetic energy of the colliding particles,
\begin{equation}
   \chi(b,E) = \pi \, - 2\, b \int_{r_m}^\infty \frac{dr}{r^2 F(r,b,E) },
    \label{eq:IntCol3}
\end{equation}
where $r_m$ is the classical returning point and $F(r,b,E) = \left( 1- \Phi(r)/E - b^2/r^2 \right)^{1/2}$. Moreover, the reduced collision integrals that we use in this work are defined as
\begin{equation}
    \Omega^{^*(l,s)}(T)  = 2 \frac{ \left( \frac{2 \, \pi \, m }{k_B} \right)^{1/2} }{ (s+1)! \left( 1- \frac{1}{2} \frac{1 + (-1)^l}{1+l} \right)} \Omega^{(l,s)}(T).
    \label{eq:IntCol4}
\end{equation}
For hard-spheres the reduced collision integrals are equal to one, and there is in the literature an extensive set of calculations for the Lennard-Jones potential \cite{klein1968tables,klein1974tables,viehland1975tables,akhmatskaya1986calculation,viehland1995high,laricchiuta2007classical}, simplified expressions can be found in \cite{meier2002computer}. In particular, we used the interpolations provided by Fokin \textit{et al.} \cite{fokin} for the evaluation of collision integrals in the LJ system.  In the case of soft-core potentials, we have implemented a FORTRAN program in order to calculate the corresponding reduced collision integrals for the parameters used in this study; the procedure follows Ref.\ \cite{smith1964automatic}, changing the interval of integration in Eq.\ \eqref{eq:IntCol1} from $(0,\infty)$ to $(0,x_{\rm max})$ to grant  convergence, where $x_{\rm max}$ corresponds to the maximum possible interaction energy at zero distance (see \cite{smith1964automatic} for more details). The values of reduced collision integrals for the Morse potential were also numerically calculated; they are in line with those reported in the literature \cite{smith1964automatic,colonna2008general,popov2013approximate}. 

\begin{figure}[t!]
    \centering
    \begin{subfigure}[b]{0.45\textwidth}
        \centering
        \includegraphics[width=\textwidth]{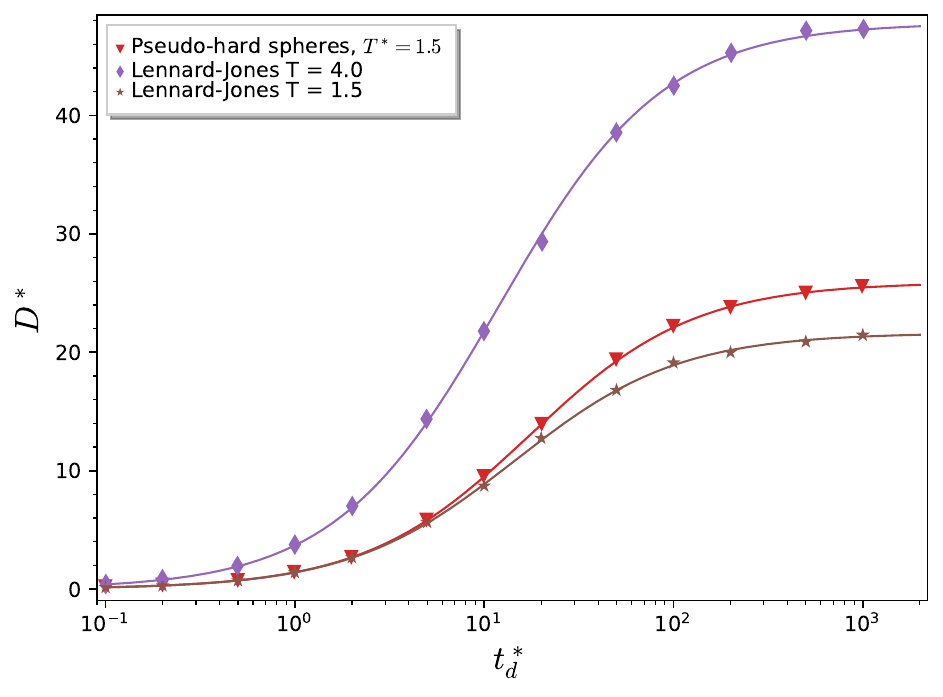}
        \caption{Hard-Core results (pseudo-hard spheres and LJ).}
        \label{fig:Diff1a}
    \end{subfigure}
    \hfill
    \begin{subfigure}[b]{0.45\textwidth}
        \centering
        \includegraphics[width=\textwidth]{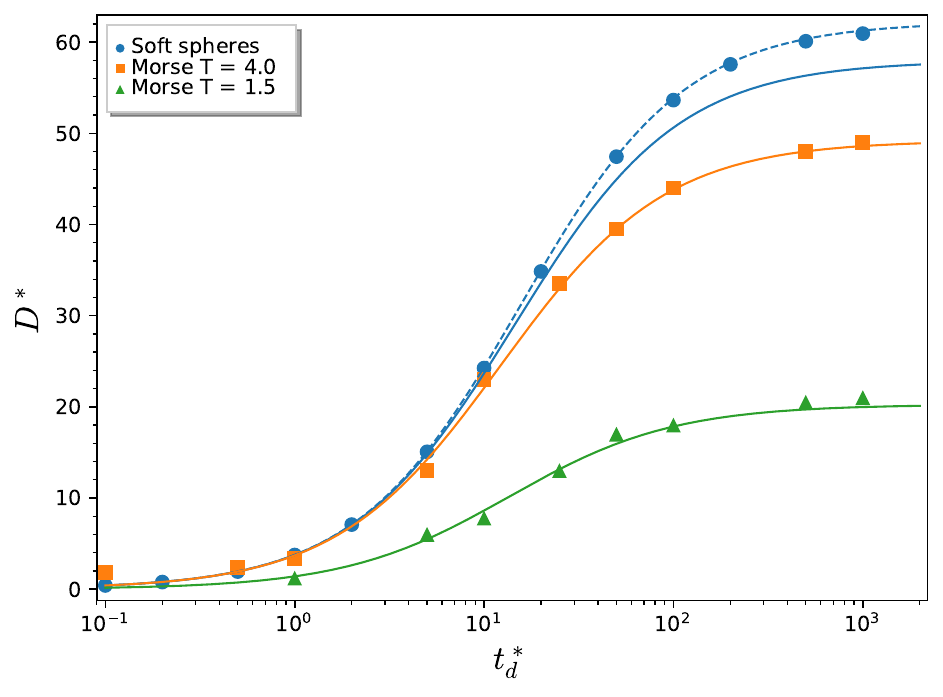}
        \caption{Soft-Core results (Morse and soft spheres).}
        \label{fig:Diff1b}
    \end{subfigure}
    \caption{Diffusion coefficient, $D^*$, against damping time, $t_d^*$ of the Langevin thermostat, in log scale for different potentials: (a) pseudo-hard spheres ($T^*=1.5$) and LJ ($T^*=1.5$ and 4), (b) soft-spheres ($T^*=4$) and Morse ($T^*=1.5$ and 4). Dots are numerical results from MD simulations and the curves are obtained form Eq.\ \eqref{e.D} with the collision integrals calculated as explained in the text. For the dotted blue curve for soft-spheres, the value of $\Omega^{*(1,1)}$ was obtained by interpolation. The concentration is $\rho^*=0.01$.}
    \label{fig:Diff1}
\end{figure}

\subsection{Simulation results}
\label{ss:sr}

The reduced diffusion coefficient, $D^*$, as a function of the reduced damping time of the Langevin thermostat, $t_d^*$, for different interaction potentials is shown in Fig.\ \ref{fig:Diff1}. The points represent  simulation values in the interval $0.1 \leq t_d^* \le 1000$ whereas the lines represent the theoretical results of Eq.\ \eqref{e.D}. The simulation results for pseudo-hard spheres and Lennard-Jones potentials in Fig.\ \ref{fig:Diff1a} were taken from Ref.\ \cite{marchioni2023dependence}, while the results for the Morse and Soft-Core potentials in Fig.\ \ref{fig:Diff1b} have been calculated for this work. In all cases, for small values of $t_d^*$, we observe a near-zero value of the diffusion coefficient, whereas for a larger damping time $D^*$ tends to a plateau with a value that corresponds to the Chapman-Enksog calculation based on the Boltzmann theory. There is a good agreement between numerical results and Eq.\ \eqref{e.D}, were the reduced collision integrals were obtained as explained in the previous section (they are well known for pseudo-hard spheres and LJ, and we calculated them for soft-spheres and Morse).

The value obtained for the collision integral for soft spheres is $\Omega^{^*(1,1)}_{T^*=4} = 0.73$ (corresponding to the solid blue curve in Fig.\ \ref{fig:Diff1b}) while, by least square fitting, we get $\Omega^{^*(1,1)}_{T^*=4} = 0.67$ (dotted blue curve). Although the discrepancy between the calculated and obtained values is not very large, this difference seems to be associated with the fact that the method used to obtain the reduced collision integrals encounters difficulties when dealing with bounded soft-core energies \cite{smith1964automatic}. On other hand, for the Morse potential, we obtained $\Omega^{^*(1,1)}_{T^*=1.5}  = 1.28$ and $\Omega^{^*(1,1)}_{T^*=4}  = 0.86$, which accurately reproduce the numerical data.

\begin{figure}[t!]
    \centering
    \begin{subfigure}[b]{0.45\textwidth}
        \centering
        \includegraphics[width=\textwidth]{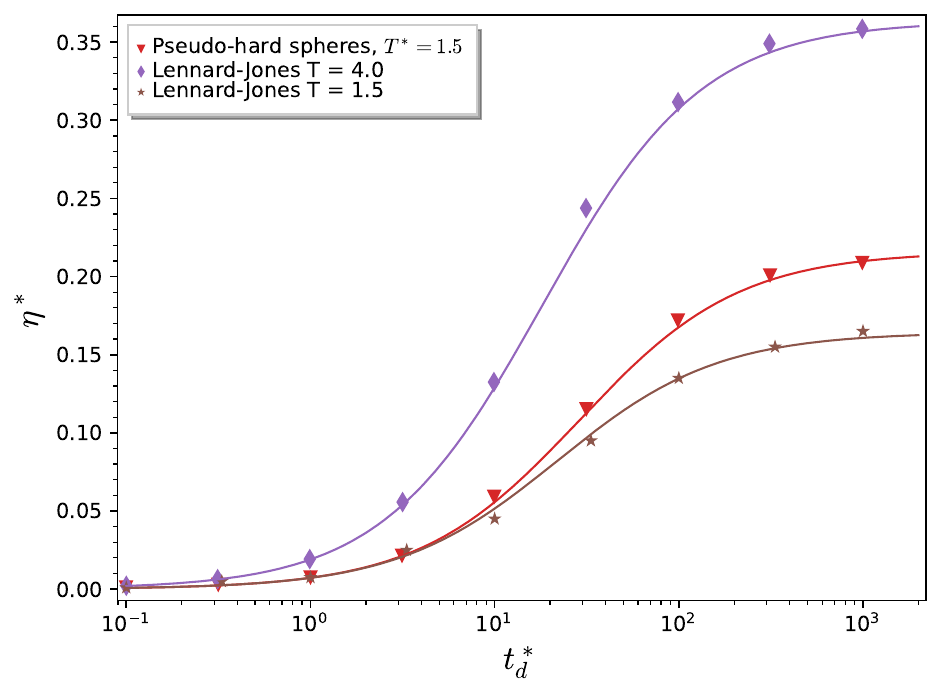}
        \caption{Hard-Core results (pseudo-hard spheres and LJ).}
        \label{fig:Visc1a}
    \end{subfigure}
    \hfill
    \begin{subfigure}[b]{0.45\textwidth}
        \centering
        \includegraphics[width=\textwidth]{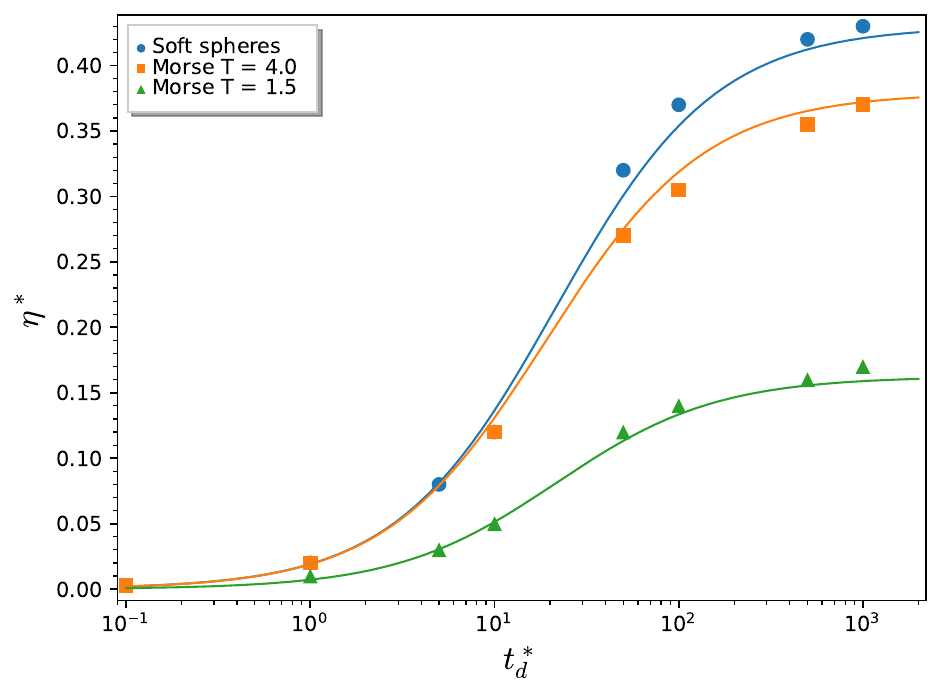}
        \caption{Soft-Core results (Morse and soft spheres).}
        \label{fig:Visc1b}
    \end{subfigure}
    \caption{Viscosity, $\eta^*$, against damping time, $t_d^*$, in log scale for different potentials: (a) pseudo-hard spheres ($T^*=1.5$) and LJ ($T^*=1.5$ and 4), (b) soft-spheres ($T^*=4$) and Morse ($T^*=1.5$ and 4). Dots are MD simulation results and curves are obtained from Eq.\ \eqref{e.e}. The concentration is $\rho^*=0.01$.}
    \label{fig:Visc1}
\end{figure}

The results for viscosity, $\eta^*$, against damping time, $t_d^*$, are presented in Fig.\ \ref{fig:Visc1}. As in the previous case, the points in the figure represent the values obtained from numerical simulations, whereas the lines correspond to the theoretical result of Eq.\ \eqref{e.e}. 
The numerical values of the viscosity for hard spheres and Lennard-Jones potentials were extracted from \cite{marchioni2024viscosity} and plotted in Fig.\ \ref{fig:Visc1a}. The results for the Morse and soft-sphere potentials in Fig.\ \ref{fig:Visc1b} were calculated for the present work, providing a broader perspective for the  understanding of the behavior of transport coefficients in the presence of noise.
The results obtained for the collision integrals for the Morse potential are $\Omega^{^*(2,2)}_{T^*=4} = 0.93$ and $\Omega^{^*(2,2)}_{T^*=1.5} = 1.33$ (corresponding to the green and orange curves in Fig.\ \ref{fig:Visc1b}) while for soft spheres it is $\Omega^{^*(2,2)}_{T^*=4} = 0.82$ (blue curve). The values for collision integrals obtained by least square fitting are $\Omega^{^*(2,2)}_{T^*=4} = 0.98$ and $\Omega^{^*(2,2)}_{T^*=1.5} = 1.30$ (Morse) and $\Omega^{^*(2,2)}_{T^*=4} = 0.80$ (soft spheres), showing a good agreement with the values calculated. This agreement strengths the robustness of the theoretical procedure leading to 
Eq.\ \eqref{e.e}.

\begin{figure}[t!]
    \centering
    \begin{subfigure}[b]{0.45\textwidth}
        \centering
        \includegraphics[width=\textwidth]{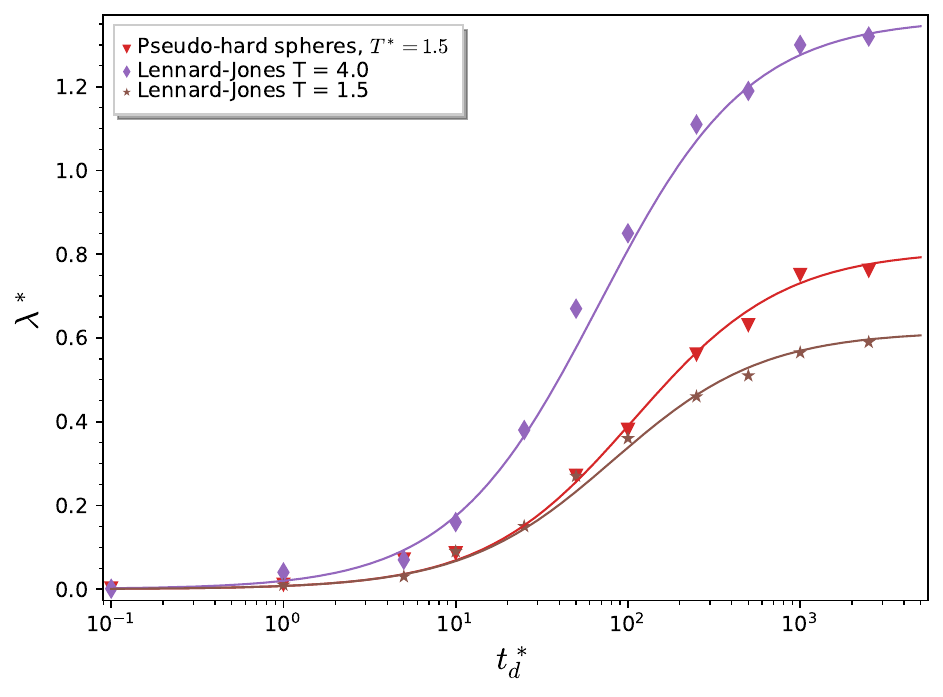}
        \caption{Hard-Core results (pseudo-hard spheres and LJ).}
        \label{fig:Therm1a}
    \end{subfigure}
    \hfill
    \begin{subfigure}[b]{0.45\textwidth}
        \centering
        \includegraphics[width=\textwidth]{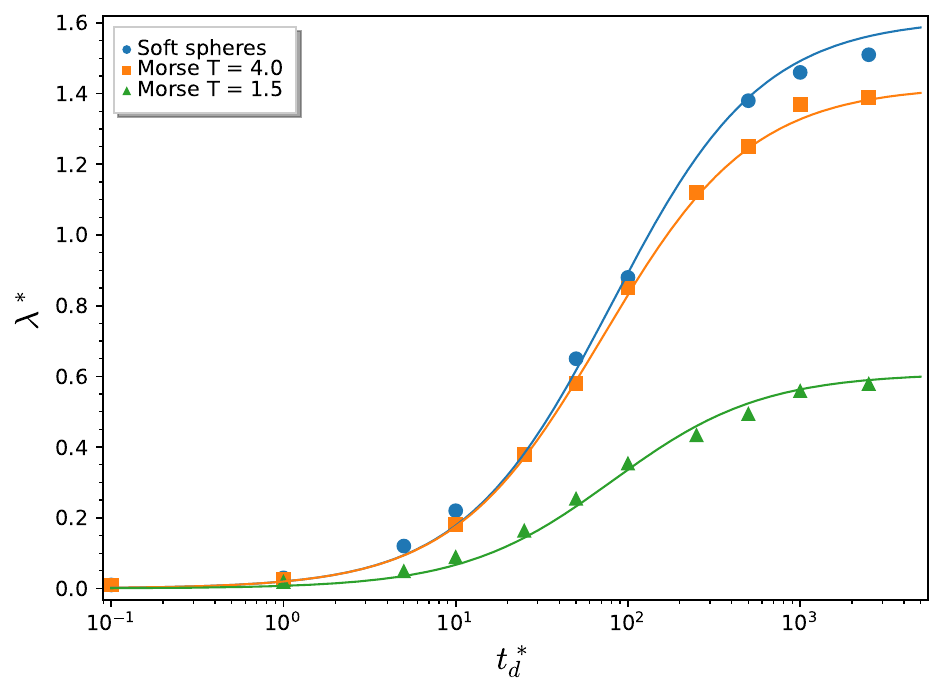}
        \caption{Soft-Core results (Morse and soft spheres).}
        \label{fig:Therm1b}
    \end{subfigure}
    \caption{Thermal conductivity, $\lambda^*$, against damping time, $t_d^*$, in log scale for different potentials: (a) pseudo-hard spheres ($T^*=1.5$) and LJ ($T^*=1.5$ and 4), (b) soft-spheres ($T^*=4$) and Morse ($T^*=1.5$ and 4). Dots are numerical results from MD simulations and the curves are obtained form Eq.\ \eqref{e.l}. The concentration is $\rho^*=0.01$.}
    \label{fig:Therm1}
\end{figure}

Finally, in Fig.\ \ref{fig:Therm1}, we present the behavior of the thermal conductivity coefficient $\lambda^*$ as a function of the damping time $t^*_d$. The results for hard-core potentials (pseudo-hard spheres and LJ) ar shown in Fig.\ \ref{fig:Therm1a} and, for soft-core potentials (soft shperes and Morse), in Fig.\ \ref{fig:Therm1b}. In the figures, we show the distinct regimes that emerge as the damping time varies, including the numerical results and curve of Eq.\ \eqref{e.l} for thermal conductivity. For pseudo-hard spheres we have that the reduced collision integrals are 1, while for LJ potential we used, as mentioned before, the analytical representations of Fokin \textit{et al.} \cite{fokin}. 
From numerical estimations of the collision integrals, we obtained $\Omega^{^*(2,2)}_{T^*=4} = 0.94$, $\Omega^{^*(2,2)}_{T^*=1.5} = 1.38$ for the Morse potential and $\Omega^{^*(2,2)}_{T^*=4} = 0.84$ for soft-spheres.
As in the previous cases, for small damping times, the thermal conductivity values are predominantly governed by the interaction with the thermal bath. In this regime, the dynamics are strongly influenced by the thermostat, which dictates the energy exchange with the environment and overshadows the role of interatomic collisions.  
In contrast, for larger damping times, the system transitions to a regime where interatomic collisions become the dominant mechanism influencing thermal conductivity. 
The interplay between these two regimes demonstrates the complex dependence of thermal transport on both external parameters and intrinsic molecular interactions.

It can be observed that in all cases, the transport coefficients for large damping times in the Morse potential are higher than those obtained for the Lennard-Jones potential, despite choosing parameters that yield a similar attractive well for both cases. This result is a consequence of the influence of the inner repulsive wall of the potentials (divergent for LJ and finite for Morse) on transport coefficients.

We observed good agreement between the theoretical description of Eqs.\ \eqref{e.D}-\eqref{e.l} and the numerical results for all the studied potentials, showing that, at least under a Langevin thermostat, the transport coefficients are predictable when the system is coupled to a thermal bath.  

\section{Conclusions}
\label{sc:cc}

In this work, we present an analysis of the effects of the Langevin thermostat on the transport coefficients for various interatomic potentials in the context of dilute gases, for which the Boltzmann theory holds. We examined a range of interaction models including both hard- and soft-core potentials. These include purely repulsive potentials, as well as those with combined attractive and repulsive components, providing a broad framework for assessing transport phenomena across different interaction types. The analysis was conducted using a combination of molecular dynamics simulations and a theoretical approach based on generalized Ohm's laws for transport processes.

The damping time associated with the Langevin thermostat serves as a control parameter, modulating the influence of the thermostat on transport coefficients. By considering a series circuit of resistances associated with interparticle collisions, described by Boltzmann's theory, and to collisions between a system particle and a reservoir particle, described by Langevin's theory, we provide an approach to describe the transport phenomena in dilute gases with a Langevin thermostat, supported by molecular dynamics simulations. These results highlight the robustness of the series circuit ansatz in capturing the interplay between the thermal bath and the intrinsic collision dynamics.

We explored a broad range of damping time scales and observed consistency between the numerical results and theoretical predictions. This consistency extends across different types of potentials, underscoring the universality of the observed behavior. Therefore, the modification of the transport coefficients of dilute gases by the Langevin thermostat can be systematically characterized and predicted.

A practical implication of these findings is that when transport coefficients are calculated for a dilute gas in contact with a Langevin thermostat, the influence of the thermostat on the results can be removed using Eq.\ \eqref{eq:rg}. 
 

\section*{Acknowledge}

This work was partially supported by Consejo Nacional de Investigaciones Cient\'ificas y T\'ecnicas (CONICET, Argentina, PUE 22920200100016CO) and Universidad Nacional de Mar del Plata (UNMDP, Argentina, 15/E1155).

\bibliographystyle{elsarticle-num}
\bibliography{main.bib}

\end{document}